\documentclass[floatfix,10pt,aps,prb,twocolumn, showpacs,a4paper, superscriptaddress]{revtex4-2}
\usepackage[T1]{fontenc}
\usepackage{amsmath}
\usepackage{amssymb}
\usepackage{graphicx}
\usepackage{mathtools}
\usepackage{bm}% bold math
\usepackage{makecell, cellspace, caption}
\usepackage{multirow}
\usepackage{array}
\usepackage{caption}
\usepackage{xcolor}
\usepackage{booktabs}
\usepackage{url}
\usepackage{hyperref}
\usepackage[normalem]{ulem} % For \sout

\usepackage{braket}
\usepackage{ragged2e}

\newcommand{\redsout}[1]{
  \begingroup
    \color{red}
    \def\UL@color{\relax}
    \sout{#1}
  \endgroup
}
\begin{document}

% Use the \preprint command to place your local institutional report
% number in the upper righthand corner of the title page in preprint mode.
% Multiple \preprint commands are allowed.
% Use the 'preprintnumbers' class option to override journal defaults
% to display numbers if necessary
%\preprint{}

%Title of paper
%\title{Monte Carlo Simulation of magnetic materilas using experimental Heisneberg exchange intercations data} % from DFT Calculation}
%\title{Standardizing Inelastic Neutron Scattering Data for Magnetic Materials and Evaluating a Quantum Correction for Classical Monte Carlo Simulations}
\title{Experimental Exchange Interaction Dataset for Magnetic Materials: Spin Waves to MC Simulations}

% repeat the \author .. \affiliation  etc. as needed
% \email, \thanks, \homepage, \altaffiliation all apply to the current
% author. Explanatory text should go in the []'s, actual e-mail
% address or url should go in the {}'s for \email and \homepage.
% Please use the appropriate macro foreach each type of information

% \affiliation command applies to all authors since the last
% \affiliation command. The \affiliation command should follow the
% other information
% \affiliation can be followed by \email, \homepage, \thanks as well.
%\email[]{Your e-mail address}
%\homepage[]{Your web page}
%\thanks{}
%\altaffiliation{}
\author{Mojtaba Alaei}
\email{m.alaei@skoltech.ru, m.alaei@iut.ac.ir}
\affiliation{Skolkovo Institute of Science and Technology, Bolshoy Boulevard 30, bld. 1, Moscow 121205, Russia}
\affiliation{Department of Physics, Isfahan University of Technology, Isfahan 84156-83111, Iran.}
\author{Zahra Mosleh}
\affiliation{Department of Physics, Isfahan University of Technology, Isfahan 84156-83111, Iran.}
\author{Nafise Rezaei}
\affiliation{Skolkovo Institute of Science and Technology, Bolshoy Boulevard 30, bld. 1, Moscow 121205, Russia}
\author{Artem R. Oganov}
\affiliation{Skolkovo Institute of Science and Technology, Bolshoy Boulevard 30, bld. 1, Moscow 121205, Russia}
%\affiliation{Materials Discovery Laboratory, Skolkovo Institute of Science and Technology, Bolshoy Boulevard 30, bld. 1, Moscow 121205, Russia}

%Collaboration name if desired (requires use of superscriptaddress
%option in \documentclass). \noaffiliation is required (may also be
%used with the \author command).
%\collaboration can be followed by \email, \homepage, \thanks as well.
%\collaboration{}
%\noaffiliation

\date{\today}

\begin{abstract}
Inelastic neutron scattering (INS) provides direct insights into microscopic magnetic interactions in crystalline materials, making it a valuable experimental technique in condensed matter physics and materials science. These interactions can be extracted by fitting spin wave dispersions to Heisenberg Hamiltonians using spin wave theory. However, such datasets are scattered across the literature and lack a standardized format, which limits their accessibility, reproducibility, and utility.
In this work, we compile and standardize exchange interaction data obtained from INS experiments on nearly 100 magnetic materials. The resulting dataset includes exchange parameters expressed in a unified Heisenberg model format, visualizations of crystal structures with annotated exchange pathways, and Monte Carlo simulation files generated using the ESpinS code. We use these experimentally derived exchange interactions to compute magnetic transition temperatures ($T_c$) via classical Monte Carlo simulations. Furthermore, we examine the impact of the $(S+1)/S$ correction in the simulations and find it improves agreement with experimental $T_c$ values in most cases.
All data and related resources are openly available through a public GitHub repository.
\end{abstract}

% insert suggested keywords - APS authors don't need to do this
%\keywords{}

%\maketitle must follow title, authors, abstract, and keywords
\maketitle

\section{Introduction}
Inelastic neutron scattering (INS) is a powerful technique for probing magnetic excitations, such as magnons—quasiparticles that represent collective spin-wave excitations in a lattice~\cite{INS1, INS2, INS3}. By measuring the energy and momentum transferred during neutron scattering events, INS provides direct insight into magnon dispersion relations. These dispersions are highly sensitive to the underlying magnetic exchange interactions, making INS an invaluable tool for characterizing magnetic systems.

Magnetic interactions in a material can be extracted by analyzing magnon excitation data through spin-wave theory,  which models magnetic excitations based on an effective spin Hamiltonian~\cite{INS1, SpinW}. However, employing full quantum spin operators in such Hamiltonians introduces significant theoretical and computational complexity. To overcome this challenge, \textit{linear spin-wave theory} (LSWT) serves as a widely used approximation method~\cite{swt1, swt2, INS1}. LSWT simplifies the problem by linearizing quantum fluctuations around an ordered magnetic ground state, expressing spin operators in terms of bosonic creation and annihilation operators. This reduces the Hamiltonian to a quadratic form, enabling an analytical solution. For example, in a simple antiferromagnetic system with only nearest-neighbor interactions, the spin Hamiltonian is given by:
$\mathcal{H} = -\sum_{\langle i,j \rangle} J\, \mathbf{S}_i \cdot \mathbf{S}_j$
where $J$ is the exchange interaction, and the summation $\langle i,j \rangle$ runs over nearest-neighbor spin pairs. Within LSWT, the magnon dispersion relation for this model simplifies to:
$E(\mathbf{k}) = -J Z S \sqrt{1 - \gamma_{\mathbf{k}}^2}$
where $Z$ is the coordination number (i.e., the number of nearest neighbors), and $\gamma_{\mathbf{k}}$ is a geometric structure factor defined as:
$\gamma_{\mathbf{k}} = \frac{1}{Z} \sum_{\boldsymbol{\delta}} e^{i \mathbf{k} \cdot \boldsymbol{\delta}}$.
Here, $\boldsymbol{\delta}$ represents the displacement vectors connecting a reference site to its nearest neighbors~\cite{swt2}. 

%\redsout{Despite the availability of theoretical tools such as LSWT for analyzing INS data, INS measurements remain limited due to the specialized and resource-intensive equipment required. Our literature review identifies only about 100 studies that report both magnon INS data and corresponding spin model mappings using LSWT. Compiling and standardizing this data into a unified framework would greatly benefit magnetism research.}
%%%%%%%%%%%%%%%%%%%%%%%%%%%%%%
Despite the availability of theoretical frameworks such as LSWT for analyzing INS data, the number of reported exchange interactions derived from INS remains limited. This is due to the specialized instrumentation, high-quality large single crystals, and significant beam time required~\cite{INS1,INS2, INS3}, making INS far less common than elastic neutron diffraction for determining magnetic structures ~\cite{Magndata, Cox1972, ENS}. Our literature survey identified only about 100 studies that report both magnon spectra from INS and corresponding spin-model parameterizations using LSWT. Establishing a unified and standardized database for such results would be a valuable resource for the magnetism community, complementing existing magnetic structure repositories.
%%%%%%%%%%%%%%%%%%%%%%%%%%%

Such a dataset could serve as a foundation for integrating future high-quality INS results and support systematic theoretical studies aimed at improving computational methods and predictive models. Additionally, establishing validation criteria is essential, as discrepancies often arise between studies on the same material, or even within a single study where multiple spin models are proposed. A fast, reliable theoretical approach such as classical Monte Carlo (MC)~\cite{MC_book} to evaluate model accuracy would greatly assist in identifying the most appropriate spin Hamiltonian

When spin-wave experimental values for Heisenberg exchange interactions are directly applied in classical MC simulations, the resulting transition temperatures $T_c$ often deviate from expectations. These discrepancies stem from the quantum effects accounted for in spin-wave theory, which are integral to deriving Heisenberg exchange parameters from neutron scattering experiments. To reconcile this mismatch in classical MC simulations, the $(S+1)/S$ correction, where $S$ is the spin magnitude of magnetic atoms (spin quantum number), has been proposed—either directly to the exchange parameters or indirectly to the $T_c$ values obtained from the simulations~\cite{Alaei2023,CMC_VS_QMC1, CMC_VS_QMC2}.  In a previous study~\cite{Alaei2023}, we used Heisenberg exchange interactions derived from INS data for 13 magnetic materials to predict their transition temperatures using classical MC simulations. By applying the $(S+1)/S$ correction to the MC results, we achieved a mean absolute percentage error (MAPE) of 8\% in the predicted transition temperatures. 

Building on our previous findings, this work aims to systematically assess the influence of the \((S+1)/S\) correction on magnetic transition temperatures across a broader range of materials. We extend our dataset to encompass 
%\redsout{approximately 73} 
72 inelastic neutron scattering (INS) studies, ensuring that all extracted exchange parameters are standardized within a consistent spin model Hamiltonian. Using classical MC simulations, we compute transition temperatures both with and without the correction. Our analysis confirms that incorporating the \((S+1)/S\) factor significantly improves the agreement between simulated and experimental \(T_c\) values. To support further research and reproducibility, the complete dataset has been made publicly available on GitHub~\cite{INS_data}.
%%%%%%%%%%%%%%%%%%%%%%%
The data is openly accessible and can be freely forked, extended, or updated with new contributions. In addition, for long-term preservation and citability, the dataset has been archived in the Zenodo repository~\cite{zenodo_ins}.
%%%%%%%%%%%%%%%%%%%%%%%%%%%

The paper is organized as follows: the Methods section outlines the methodology employed in this study, while the Results section examines the use of classical MC simulations both for predicting magnetic transition temperatures and for identifying inaccuracies in reported exchange parameters derived from INS data. Finally, the Conclusions section summarizes our findings and discusses their broader implications.

\section{Methods}

\subsection{Data gathering and standardizing}

We attempted to identify as many research papers as possible that present INS data (magnetic excitations) analyzed using spin wave theory to extract exchange interactions.
A review of the literature shows that the Heisenberg term in spin model Hamiltonians typically appears in one of the following forms:

\begin{eqnarray}
H_{Heis.}&=& -\sum_{i,j} J_{i,j}  \mathbf{S}_i \cdot \mathbf{S}_j  \nonumber \\ 
H_{Heis}&=& -\frac{1}{2}  \sum_{i,j} J_{i,j}  \mathbf{S}_i \cdot \mathbf{S}_j \nonumber \\ 
H_{Heis}&=& -\sum_{<i,j>} J_{i,j}  \mathbf{S}_i \cdot \mathbf{S}_j  \nonumber \\
H_{Heis}&=& -2\sum_{<i,j>} J_{i,j}  \mathbf{S}_i \cdot \mathbf{S}_j   \nonumber
\end{eqnarray}

\begin{figure*}
\centering
\includegraphics[width=1.0\textwidth]{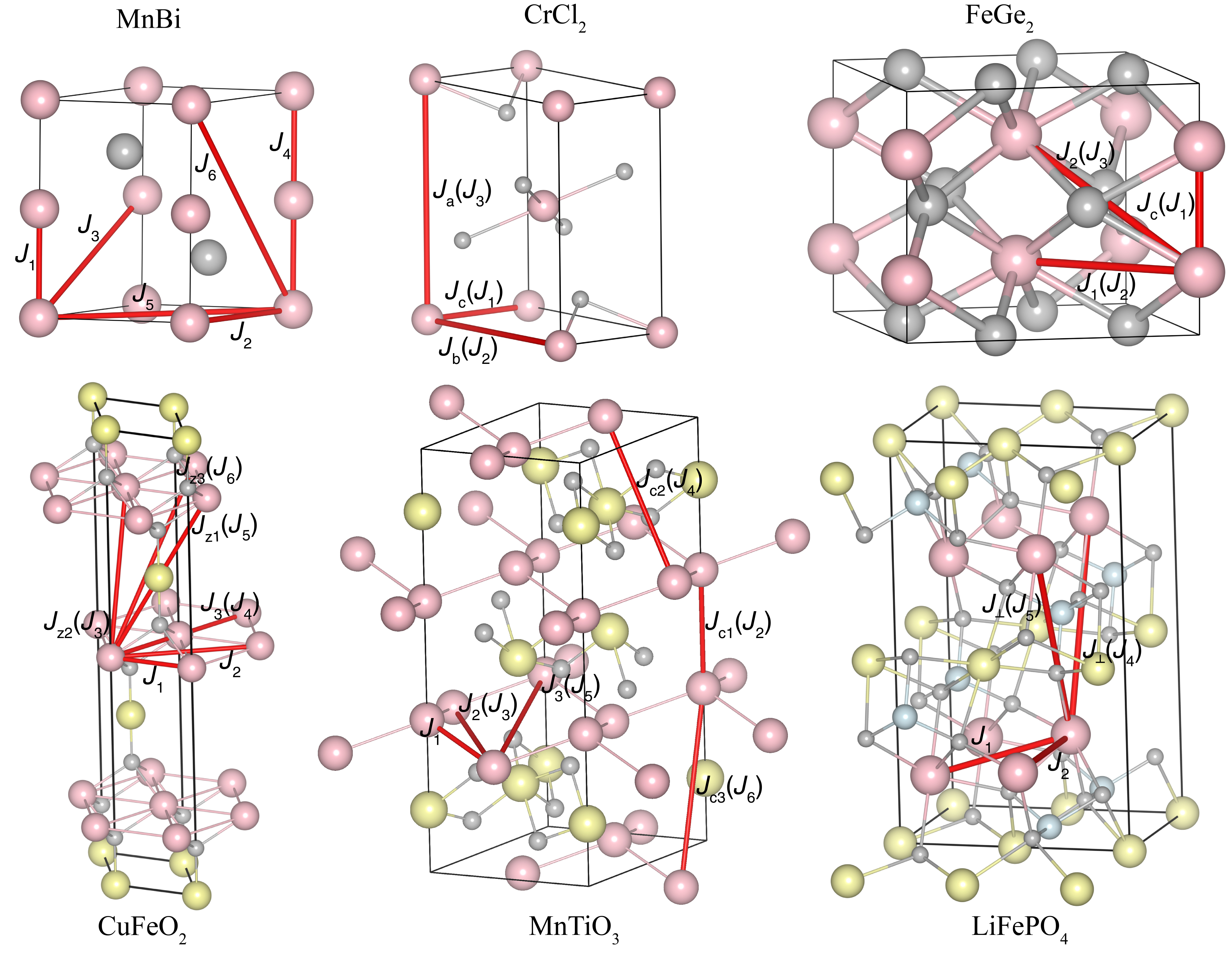}
\caption{\justifying
Illustrative images of sample crystal structures from our database. Each image highlights the exchange interactions and their labeling, as presented in the corresponding INS paper. The labels in parentheses indicate the ranking of exchange interactions based on nearest-neighbor distances (e.g., $J_1$ corresponds to the first nearest neighbor,$J_2$ to the second nearest neighbor, and so on).}
\label{fig:samples}
\end{figure*}
 
One of the key challenges in standardizing exchange interactions is the ambiguity in the choice of spin Hamiltonian conventions, particularly regarding the double counting of pairwise interactions. For example, models that use  $-\frac{1}{2} \sum_{i,j} J_{i,j} \, \mathbf{S}_i \cdot \mathbf{S}_j$ 
or 
$-\sum_{\langle i,j \rangle} J_{i,j} \, \mathbf{S}_i \cdot \mathbf{S}_j$  
are designed so that each interaction between spins at sites \( i \) and \( j \) is counted only once.  
In contrast, models that adopt  
$-\sum_{i,j} J_{i,j} \, \mathbf{S}_i \cdot \mathbf{S}_j$ 
or   $ -2\sum_{\langle i,j \rangle} J_{i,j} \, \mathbf{S}_i \cdot \mathbf{S}_j$  
count each pairwise interaction twice.
A major source of confusion arises when some studies claim to use the convention  
$-\sum_{i,j} J_{i,j} \, \mathbf{S}_i \cdot \mathbf{S}_j$, 
yet a detailed analysis—often involving comparisons with other works—reveals that their reported exchange constants have already been corrected to avoid double counting. In practice, these studies effectively use the convention  
$-\frac{1}{2} \sum_{i,j} J_{i,j} \, \mathbf{S}_i \cdot \mathbf{S}_j$,  
as seen, for example, in Refs.~\cite{YbMnSb2, YbMnBi2, MnBi2Te4_prl, MnF2_Yamani, INS3}.

A particularly clear example of this mismatch between the stated Hamiltonian and the magnon dispersion derived from spin wave theory is found in the altermagnet MnF$_2$~\cite{MnF2_Yamani, INS3}. This compound has been extensively studied in INS experiments~\cite{MnF2, MnF2_OKAZAKI19649, MnF2_Yamani, MnF2_Takashi} and is widely regarded as a textbook prototype of antiferromagnetic order~\cite{INS3, MnF2_book1, MnF2_book2}.

Additionally, some papers use a positive sign in front of the summation instead of a negative one. In such cases, the interpretation of \( J_{i,j} \) is reversed: a negative value of \( J_{i,j} \) indicates a ferromagnetic interaction, while a positive value corresponds to an antiferromagnetic interaction.

For magnetic anisotropy, the following forms are commonly considered in the literature:
\begin{eqnarray} \nonumber 
H_{\text{Ani}} &=& -D\sum_{i}  S_{i,z}^2  \nonumber \\ 
H_{\text{Ani}} &=&  -D_x\sum_{i}  S_{i,x}^2 - D_z\sum_{i}  S_{i,z}^2, \nonumber 
\end{eqnarray} 
where $D$, $D_x$, and $D_z$ represent the anisotropy strengths along specific directions. 

To ensure consistency, we express the exchange and magnetic anisotropy interactions in the following standardized form:

\begin{equation}\label{Eq1}
	H=-\frac{1}{2}  \sum_{i,j} \tilde{J}_{ij} \hat {\mathbf{S}}_i \cdot \hat {\mathbf{S}}_j   +  \sum_i  \hat {{\mathbf S}}_i  \tilde{{\mathbf A}} \hat {{\mathbf S}}_i.
\end{equation}

\noindent
Here, $\tilde{J}_{ij}$ (obtained by rescaling ${J}_{ij}$ by $S^2$) represents the Heisenberg exchange interaction strength between sites $i$ and $j$, while $\hat {\mathbf{S}}_i$ and $\hat {\mathbf{S}}_j$ denote unit vectors indicating the magnetic moment direction at lattice sites $i$ and $j$. The matrix $\tilde{{\mathbf A}}$ characterizes the anisotropy. 

To avoid additional complexity, we do not consider the Dzyaloshinskii-Moriya interaction (DMI) in this work. Therefore, we select data where DMI is either not reported or is negligible compared to the dominant Heisenberg exchange interaction.  

Standardization also requires knowledge of the spin quantum number $S$, as it is necessary for rescaling $J_{ij}$ to $\tilde{J}_{ij}$. Consequently, we exclude studies where the value of $S$ is ambiguous. When available, we adopt the value of $S$ reported in INS  studies, as it is typically consistent with the experimentally measured magnetic moment.

In the literature, interaction strengths are expressed in various units, including THz, meV, Kelvin, and $\text{cm}^{-1}$. More recent studies tend to report magnetic interactions in meV. Therefore, we standardize all interaction strengths in meV for consistency.

\subsection{Visualization and MC simulations}

For each compound, provide an image of the crystal structure including illustration for $J_{i,j}$
between first, second ,.., {\it nth} nearest neighbors of magnetic atoms indicating by $J_1$, $J_2$ and $J_n$ respectively (Fig.~\ref{fig:samples}).
In some papers, researchers prefer to indicate Heisenberg interactions $J_{i,j}$ by indices of lattice vectors 
(e.g, $J_{c}$ indicates exchange interaction between magnetic atoms along lattice c axis).
In such cases, we indicate our standard naming of exchange interactions inside parentheses (Fig.~\ref{fig:samples}).

We use VESTA~\cite{Vesta} for structure visualization.  
For each compound, we provide structure files in VESTA format, which not only allow researchers to visualize the atomic positions and lattice vectors but also highlight Heisenberg exchange interactions using thick red lines.

We use the ESpinS~\cite{ESpinS} code for MC simulations.  
For each compound, we provide all the necessary input and output files for ESpinS code~\cite{INS_data}.  
Our simulations indicate that using approximately 2000 lattice sites is generally sufficient to predict the transition temperature.  
Therefore, for MC simulations, we use supercells that contain a minimum of 2000 sites.
To determine the transition temperature, we identify the peak of the magnetic specific heat, given by  $C_M = \frac{\braket{E^2} - \braket{E}^2}{Nk_B T^2}$.    
In cases where the exact peak position is unclear, we also analyze the fourth-order energy cumulant, defined as    
$U_E = 1 - \frac{1}{3} \frac{\braket{E^4}}{\braket{E^2}^2}$.  

\section{Data Records}

For each material in the GitHub database~\cite{INS_data},  
there is a dedicated directory named after the compound. Each directory contains the following information:  

\begin{itemize}
    \item \textbf{Structure file}: A VESTA format file with thick lines indicating exchange interactions.  
    \item \textbf{Visualization}: A JPEG image showing atomic positions and exchange interactions, as depicted in Fig.~\ref{fig:samples}. This image is displayed online for each compound.  
    \item \textbf{Data tables}: Tables listing exchange interactions (in meV), spin quantum number \(S\), and both experimental and MC transition temperatures.  
    \item \textbf{Reference links}: Sources for INS data and transition temperature values.  
    \item \textbf{MC simulation files}: A \texttt{MC} directory containing all input and output files from MC simulations performed using \texttt{ESpinS}.  
\end{itemize}

\section{Technical Validation}

\subsection{Predicting transition temperature using Monte Carlo}
The quantum Monte Carlo (QMC) method is among the most accurate approaches for determining the transition temperatures of magnetic materials. However, its applicability is often limited by the fermion sign problem, especially in frustrated systems ~\cite{QMC}. Additionally, accounting for finite-size effects in QMC simulations is highly computationally demanding.
 
Other methods, such as mean field theory (MFT)~\cite{MFT_Johnston} and the random phase approximation (RPA)~\cite{RPA1, MC_RPA_MFT1, MC_RPA_MFT2}, are also used. MFT tends to overestimate the transition temperature compared to experimental results, while RPA improves upon MFT and yields more accurate predictions~\cite{MC_RPA_MFT1, MC_RPA_MFT2, MC_RPA_MFT3, MC_RPA_MFT4}. However, applying RPA requires specific techniques and details, such as the magnetic ordering vector, which limits its use to researchers with sufficient theoretical knowledge~\cite{MC_RPA_MFT3}.  

Classical MC can be a useful tool for providing a rough estimate of the transition temperature in magnetic materials. However, when using INS  data fitted with spin-wave theory, classical MC simulations tend to underestimate the transition temperature~\cite{Alaei2023}. 
This discrepancy casts doubt on the reliability of the INS data~\cite{MC_spinwave}.
Since spin-wave theory incorporates quantum mechanical effects in fitting inelastic neutron scattering (INS) data, including the factor \((S + 1)/S\) is essential for obtaining accurate results from classical Monte Carlo (MC) simulations. This factor arises from comparing the quantum and classical expressions for the expectation value of the squared spin operator. In the quantum case, \(\langle \mathbf{S}^2 \rangle\) is proportional to \(S(S + 1)\), whereas in the classical limit it scales as \(S^2\). To reconcile these two approaches, either the exchange parameters used in classical MC simulations or the transition temperatures obtained from classical MC must be rescaled by the factor \((S + 1)/S\)~\cite{Alaei2023}.

Figure~\ref{fig:T} shows the transition temperatures obtained from MC simulations (\(T_{\mathrm{MC}}\)) for all 72 compounds, revealing a significant deviation from the experimental transition temperatures. However, applying the \((S+1)/S\) correction:  
\begin{equation}  
T_{\mathrm{MC}}^{*} = \frac{S+1}{S} T_{\mathrm{MC}}  
\end{equation}  
significantly improves the accuracy of transition temperature predictions. 

The MAPE for $T_{\mathrm{MC}}^*$ is 9.6\%, whereas for $T_{\mathrm{MC}}$, the error is much greater, equal to 35.1\%. %This result is consistent with our previous study \cite{Alaei2023}, which was based on only 13 compounds.
For 36\% of the compounds, the absolute percentage error (APE) is less than 5\%. In  32\% of the cases, APE falls between 5\% and 10\%, while for 11\% of the compounds, it ranges from 10\% to 15\%. Only 21\% of the compounds exhibit an APE greater than 15\%. 
The detailed data from the MC simulations shown in Fig.~\ref{fig:T} are provided in Appendix~\ref{app:A}.

%\redsout{It is important to note that exchange interactions derived from INS data are subject to uncertainties due to their reliance on pure fitting to LSWT. Additionally, exchange interactions can vary with temperature in practice
%~\cite{J_vs_temp1,J_vs_temp2, J_vs_temp3}. 
%These factors collectively impact the accuracy of predicting the transition temperature using simulation methods such as MC.}
It is important to note that exchange interactions derived from INS data are subject to several approximations. First, LSWT itself is an approximation to spin-wave theory when fitting INS data to a spin Hamiltonian. Second, in practice, only the dominant exchange terms are typically included in the spin Hamiltonian, while weaker interactions (such as single-ion anisotropy or longer-range couplings) are often neglected. Finally, exchange parameters are not strictly constant but may vary with temperature in practice~\cite{J_vs_temp1,J_vs_temp2,J_vs_temp3}. Another source of error arises from the classical nature of Monte Carlo simulations: while classical spin models become exact in the large-S limit, quantum effects can play a role for small spin values. Nevertheless, previous studies have shown that for $S\geq1$ the difference between quantum and classical Monte Carlo results is usually minor, and even for $S=\frac{1}{2}$ the deviations remain small~\cite{CMC_VS_QMC1,CMC_VS_QMC2}. Finite-size effects in the simulations can also lead to systematic shifts in the transition temperature. Taken together, these experimental and theoretical limitations account for the observed deviations between our predicted and experimental transition temperatures.

As a result, calculating an exact transition temperature is not solely dependent on theoretical methods; it also relies on the accuracy of experimentally derived exchange interactions. This means that even with a perfect theoretical approach, expecting a transition temperature that exactly matches experimental results is unrealistic.

\subsection{Refining Spin Hamiltonians Using Monte Carlo Results}

\begin{figure}
    \centering
    \includegraphics[width=0.9\linewidth]{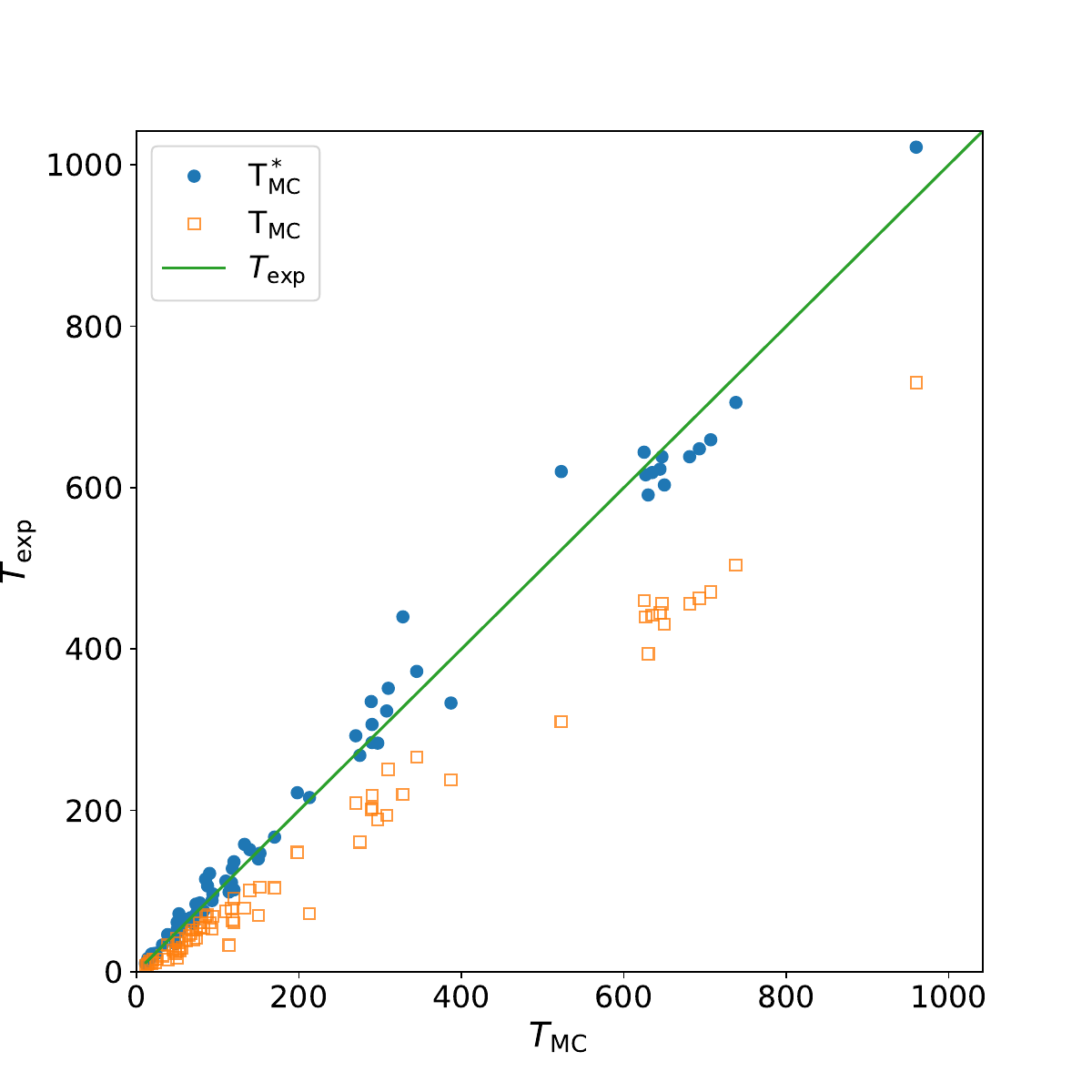}
    \caption{ \justifying
Results from MC simulations for 72 compounds.  
$T_{\mathrm{MC}}$ represents the transition temperature obtained from MC simulations using exchange parameters derived from INS  data fitted with spin-wave theory.  
$T_{\mathrm{MC}}^{*}$ denotes the quantum-corrected transition temperature, obtained by applying the correction factor $(S+1)/S$ to $T_{\mathrm{MC}}$. 
All data points shown in this diagram are provided in Appendix~\ref{app:A}.}
    \label{fig:T}
\end{figure}

From the plot (Fig.~\ref{fig:T}), we can establish two key observations:  

\begin{enumerate}  
    \item The transition temperature obtained from classical MC simulations ($T_{\mathrm{MC}}$) is always lower than the experimental transition temperature ($T_{\mathrm{Exp.}}$), i.e., $T_{\mathrm{MC}} < T_{\mathrm{Exp.}}$.  
    \item The quantum correction factor \((S+1)/S\) accurately accounts for this discrepancy in most cases.  
\end{enumerate}  

These observations provide useful criteria for assessing the accuracy of exchange parameters derived from INS data using spin-wave theory.
In the following, we discuss several cases to illustrate how these criteria can be used to identify better models and filter out unreliable data.

NiO is one of the most extensively studied antiferromagnetic compounds, with a transition temperature of 523 K and a quantum spin number $S = 1$. However, only a single INS dataset exists for it~\cite{NiO}. In the INS study, two different Heisenberg models were used to fit the data: a $J_1-J_2$ model (including only first- and second-nearest-neighbor exchange interactions) and a $J_1-J_4$ model (including interactions up to the fourth nearest neighbor).  

Using the $J_1-J_2$ model, the MC simulation yields $T_{\mathrm{MC}} = 310$ K and $T_{\mathrm{MC}}^* = 620$ K. In contrast, the \(J_1\)-\(J_4\) model produces $T_{\mathrm{MC}} = 304$ K and $T_{\mathrm{MC}}^* = 608$ K, suggesting that the $J_1-J_4$ model provides a more accurate estimate of the transition temperature.  

\begin{figure}
\centering
\includegraphics[width=0.9\linewidth]{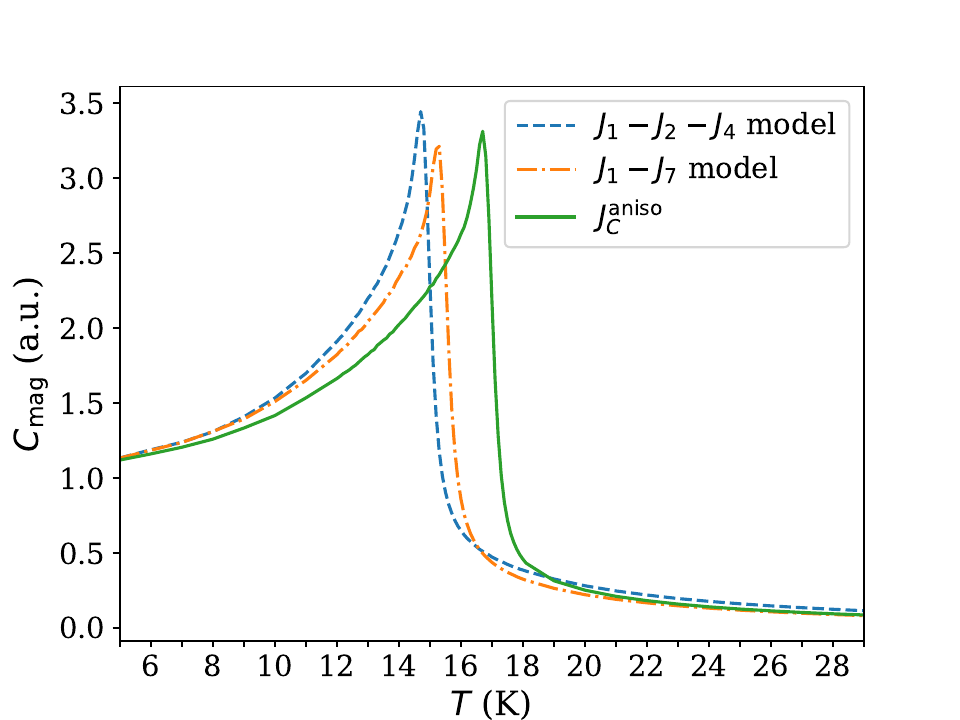}
\caption{ \justifying
Magnetic specific heat calculated from MC simulations for three different spin models of MnBi$_2$Te$_4$, based on parameters obtained from INS data. The $J_1$–$J_2$–$J_4$ model was introduced in Ref.~\cite{MnBi2Te4_prl}, while the $J_1$–$J_7$ and $J_c^{\mathrm{aniso}}$ models were proposed in Ref.~\cite{MnBi2Te4}. The peak of the specific heat defines the MC transition temperature, $T_{\mathrm{MC}}$. The values of $T_{\mathrm{MC}}$ for the $J_1$–$J_2$–$J_4$, $J_1$–$J_7$, and $J_c^{\mathrm{aniso}}$ models are 14.7~K, 15.3~K, and 16.7~K, respectively.}
\label{fig:Cm}
\end{figure}
Another example is MnBi$_2$Te$_4$, for which two successive research papers were published by overlapping groups of authors~\cite{MnBi2Te4_prl, MnBi2Te4}. This compound has an experimental magnetic transition temperature of 24~K and a spin quantum number of $S = 5/2$. All spin models considered in these studies include intra-layer exchange interactions ($J_1, J_2, \dots$) and an inter-layer exchange interaction ($J_c$).

In the first study~\cite{MnBi2Te4_prl}, the authors introduced the $J_1$--$J_2$--$J_4$ model, which resulted in a MC transition temperature $T_{\mathrm{MC}} = 14.7$~K and a corrected transition temperature $T_{\mathrm{MC}}^* = 20.58$~K (See Fig.~\ref{fig:Cm}). 
In the more recent work~\cite{MnBi2Te4}, the model was extended to include exchange interactions up to the 7\textsuperscript{th} nearest neighbor ($J_1$--$J_7$), and an alternative model incorporating anisotropic inter-layer exchange ($J_c^{\mathrm{aniso}}$) was also proposed.

MC simulations using the $J_1$--$J_7$ model yield $T_{\mathrm{MC}} = 15.3$~K and $T_{\mathrm{MC}}^* = 21.42$~K (See Fig.~\ref{fig:Cm}). For the $J_c^{\mathrm{aniso}}$ model, the results improve further to $T_{\mathrm{MC}} = 16.7$~K and $T_{\mathrm{MC}}^* = 23.38$~K, suggesting that the $J_c^{\mathrm{aniso}}$ model provides the best agreement with the experimental data.

During our investigation, we found instances where the transition temperature obtained from Monte Carlo simulations ($T_{\mathrm{MC}}$) exceeded the experimental value, and other cases where the quantum-corrected temperature ($T_{\mathrm{MC}}^*$) was significantly lower. Such discrepancies often indicate problems with the INS data or the fitting procedure. These issues may stem from an incorrectly defined Hamiltonian—such as a missing factor of $1/2$—poor data fitting, or an inaccurate estimate of the spin quantum number $S$.  
For example, using the INS-derived exchange interactions for LiNiPO$_4$~\cite{LiNiPO4}, LiCoPO$_4$~\cite{LiCoPO4}, and LiMnPO$_4$~\cite{LiMnPO4}, the MC-calculated transition temperatures ($T_{\mathrm{MC}}$) are  overestimated. This points to a mismatch between the spin Hamiltonian and the spin-wave formalism, likely caused by an omitted factor of $1/2$ in the Hamiltonian. A closer examination of the spin-wave equations used in these studies supports this conclusion.

%\section{Conclusions}
Through analysis of more than 100 INS  studies that derived exchange interactions from spin-wave theory, we identified 72 compounds with reliable data and compiled them into a standardized database. This database includes exchange interaction parameters, spin quantum numbers, reference sources, MC simulation results, structure files, and visualized structures.  
Using this dataset, we investigated the accuracy of transition temperature predictions by applying the \((S+1)/S\) correction to classical MC results. Our findings show that, in most cases, the corrected results align more closely with experimental values, achieving a mean absolute percentage error (MAPE) of 9\%.  
We propose classical MC simulations and the \((S+1)/S\) correction as useful tools for refining INS data and identifying potential inconsistencies.
%%%%%%%%%%%%%%%%%%%%%%%%%%%%%
\subsection{Comparison with theories}
\begin{figure}
\centering
\includegraphics[width=0.90\linewidth]{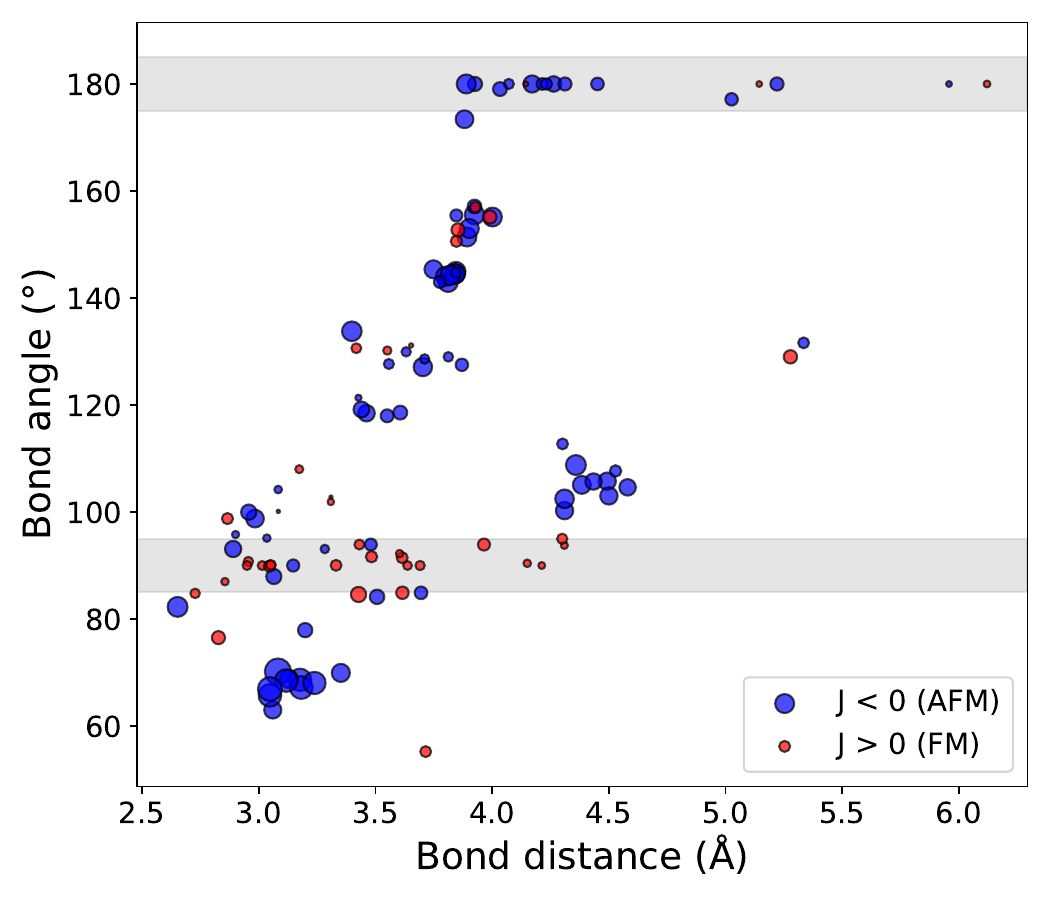}
\caption{ \justifying
The diagram illustrates the exchange interaction as a function of the bond angle (M$_1$–L–M$_2$) between magnetic atoms (M$_1$, M$_2$) and the ligand (L), and the interatomic distance (M$_1$–M$_2$). Red and blue circles represent ferromagnetic and antiferromagnetic interactions, respectively. The circle radius is proportional to the square root of the interaction strength, i.e., $\sqrt{|J|}$.}
\label{fig:J_vs_ang}
\end{figure}

To provide further insight into the data, we analyze the role of the local atomic environment in determining the exchange interactions. For this purpose, we focus on the bond angle (M$_1$–L–M$_2$) between two magnetic atoms (M$_1$, M$_2$) bridged by a ligand (L), together with the distance between the magnetic atoms (M$_1$–M$_2$). To ensure that the bond angle is well defined, we restrict the analysis to cases where the two magnetic atoms are symmetrically coordinated by the ligand (M$_1$–L = M$_2$–L). Exchange pathways involving more than one intermediate atom (e.g., M$_1$–L–M$_2$–L–M$_3$ or M$_1$–L$_1$–L$_2$–M$_2$) are excluded from the analysis, since the bond angle cannot be uniquely defined in such cases.
Figure ~\ref{fig:J_vs_ang} presents the resulting diagram of exchange interactions as a function of bond angle and interatomic distance. The overall trend is broadly consistent with one of the Goodenough–Kanamori–Anderson (GKA) rules~\cite{GKA1,GKA2,GKA3}: bond angles close to $180^\circ$ generally favor antiferromagnetic interactions ($J < 0$), while bond angles near $90^\circ$ more often lead to ferromagnetic interactions ($J > 0$). It is also important to note that the GKA rules further emphasize the dependence of exchange on the d-orbital filling of the magnetic ions.
However, INS data, together with theoretical calculations, clearly demonstrate that the GKA rules are not universally valid. A representative example is KCuF$_3$, where two Cu–F–Cu pathways both have bond angles of $\sim 180^\circ$ (corresponding to Cu–Cu distances of 3.92 \AA ~and 4.14 \AA), yet one interaction is antiferromagnetic while the other is ferromagnetic 
%[INS_ref, Theory_ref]
. This shows that while the GKA rules provide valuable qualitative guidance, the actual exchange interactions also depend sensitively on factors such as orbital hybridization, covalency, and details of the crystal structure.

\begin{figure}
\centering
\includegraphics[width=0.90\linewidth]{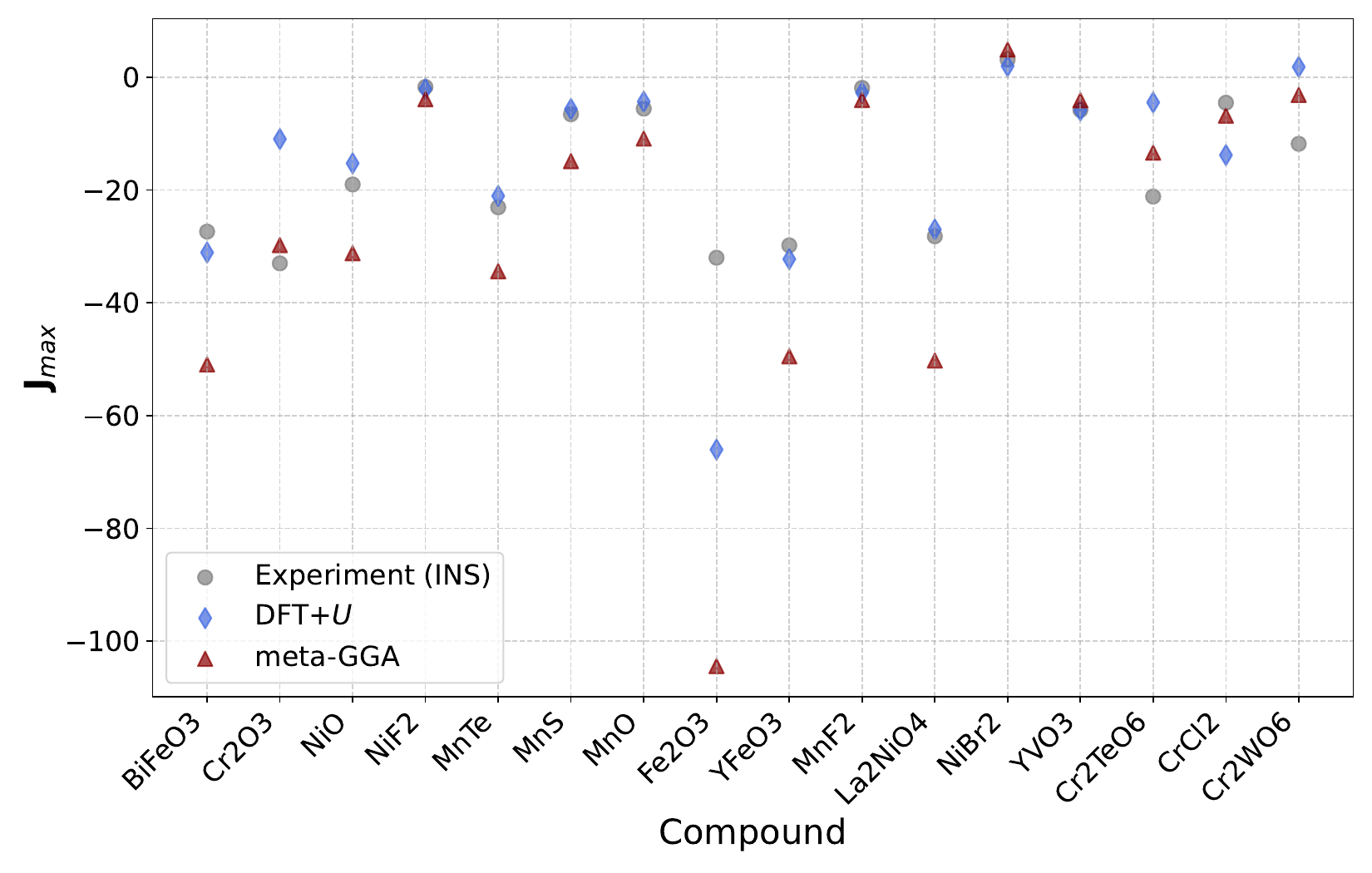}
\caption{ \justifying
Maximum exchange interaction for 13 compounds, as determined from INS measurements, compared with values obtained from DFT+$U$~\cite{Mosleh} and meta-GGA r$^2$SCAN~\cite{Nrezaei}.}
\label{fig:Js}
\end{figure}

Comparisons between INS-derived exchange parameters and those obtained from density functional theory (DFT) are highly valuable, but they must be treated with care, as DFT results can vary significantly depending on the choice of basis set, exchange–correlation functional, and other methodological details. To ensure consistency and avoid ambiguities that might arise from mixing results obtained with different approaches (e.g., those available in public databases), we restrict our comparison to two of our previous systematic studies, in which a uniform computational strategy was applied across all compounds.
In the first study~\cite{Mosleh}, we employed DFT+$U$~\cite{LDAU} with the Hubbard parameter determined self-consistently using linear response theory, while in the second study~\cite{Nrezaei} we used the meta-GGA r$^2$SCAN~\cite{r2SCAN} functional to calculate exchange interactions. Thirteen compounds are common to both of these earlier works and the present INS dataset. For these systems, we compare the maximum exchange interaction obtained from INS, DFT+$U$, and meta-GGA r$^2$SCAN (Figure ~\ref{fig:Js}).
The results reveal a clear trend: DFT+$U$ tends to underestimate the exchange interaction strength relative to INS, whereas meta-GGA r$^2$SCAN generally overestimates it. These comparisons, however, should be interpreted with caution. In our earlier DFT studies, the mapping was carried out onto a classical Heisenberg Hamiltonian, while in the present INS analysis the mapping is performed within LSWT, which explicitly incorporates quantum effects. This methodological distinction likely contributes to the observed discrepancies between theory and experiment.

%%%%%%%%%%%%%%%%%%%%%%%%%%%%%%%%
\section{Author Contributions}
Mojtaba Alaei conducted and supervised the project, performed part of the Monte Carlo (MC) simulations, prepared data for the GitHub repository, and wrote the initial draft of the manuscript. Zahra Mosleh gathered data, prepared inputs for the MC simulations, and contributed to structural visualization. Nafise Rezaei carried out MC simulations, contributed to structural visualization, and participated in manuscript writing. Artem R. Oganov provided intellectual input and financial support. All authors reviewed and edited the manuscript.

\section{Competing interests}
The authors declare no competing interests
\section{Code availability}
This study did not involve the development of new code.
\begin{acknowledgments}
M.~A.\ acknowledges financial support from the Iran National Science Foundation (INSF) under Project No.~4002648. N.~R.\ and A.~R.~O.\ acknowledge support from the Russian Science Foundation under Grant No.~19-72-30043.
\end{acknowledgments}

% Switches to one column mode
%\clearpage
%\newpage
\begin{widetext}
\appendix
%\documentclass[aps,prl,preprint,twocolumn,groupedaddress]{revtex4-2}
%\documentclass[floatfix,10pt,aps,prb,showpacs,a4paper]{revtex4-2}

%\begin{document}
\setcounter{table}{0}
\renewcommand{\thetable}{A\arabic{table}}

\section{Transition temperature} 
\label{app:A}
In this appendix, we provide all the data required to plot Figure~\ref{fig:T}.  
Table~\ref{tab:temp} summarizes the experimental transition temperatures ($T_{\mathrm{Exp.}}$), 
Monte Carlo (MC) transition temperatures ($T_{\mathrm{MC}}$), and quantum-corrected MC transition 
temperatures ($T_{\mathrm{MC}^{*}}$). For each compound, we include references to the 
inelastic neutron scattering (INS) data along with the corresponding experimental 
transition temperatures. Additional details, such as the input and output files 
from the MC simulations, are available in the GitHub repository~\cite{INS_data}
\begin{table}[h]
	%\centering
	\caption{\justifying
    Transition temperatures obtained from MC simulations. $T_{\mathrm{MC}}$ was determined through MC simulations. $T_{\mathrm{MC}}^*$ was then obtained by multiplying the MC results with the ($\frac{S+1}{S}$) factor. Additionally, $T_{\mathrm{Exp.}}$ represents the experimentally reported temperature. $S$ represents the spin of the magnetic ion, and Error denotes the percentage difference between $T_{\mathrm{MC}}^*$ and $T_{\mathrm{Exp.}}$.}
	\label{tab:temp}
    \noindent
	\begin{minipage}[t]{0.42\textwidth}
		%\centering
		\begin{tabular}[t]{lccccc}
			\hline
			& \makecell{$T_{\mathrm{Exp.}}$(K)} &  \makecell{$T_{\mathrm{MC}}$(K)}  & \makecell{$T_{\mathrm{MC}}^*$(K)}&  \makecell{S} &  \makecell{Error(\%)}\\
			\hline
			Ba$_2$NiWO$_6$\cite{Ba2NiWO6}            &  $48$  \cite{Ba2NiWO6}     &  $23$   &  $46$ &  $1$ &  $4.2$  \\
			BaMn$_2$As$_2$\cite{BaMn2As2}  &  $625$\cite{BaMn2As2}       &  $460$   &  $644$ &  $5/2$ &  $3.0$ \\
			BaMn$_2$Bi$_2$\cite{BaMn2Bi2}&  $387.2$\cite{BaMn2Bi2}      &  $238$   &  $333.2$ &  $5/2$ &  $13.9$     \\
			BaNi$_2$As$_2$O$_8$\cite{BaNi2As2O8}  &  $18.5$\cite{BaNi2As2O8_T}      &  $11 $   &  $22$ &  $1$ &  $18.9$     \\
			Bi$_2$CuO$_4$\cite{Bi2CuO4}  &  $50$ \cite{Bi2CuO4}     &  $17.5$   &  $52.5$ &  $1/2$ &  $5$     \\
			BiFeO$_3$\cite{BiFeO3}  &  $650$\cite{BiFeO3}      &  $431$   &  $603.4$ &  $5/2$ &  $7.2$     \\
			Ca$_3$Ru$_2$O$_7$\cite{Ca3Ru2O7}  &  $56$\cite{Ca3Ru2O7}      &  $30$   &  $60$ &  $1$ &  $7.1$     \\
			CaMn$_2$Sb$_2$\cite{CaMn2Sb2}  &  $85$\cite{CaMn2Sb2}      &  $69$   &  $115$ &  $3/2$ &  $35.3$     \\
			CaMn$_7$O$_{12}$\cite{CaMn7O12}  &  $90$\cite{CaMn7O12}      &  $61$   &  $122$ &  $1$ &  $35.5$     \\
			CaMnBi$_2$\cite{CaMnBi2}  &  $270$\cite{CaMnBi2}      &  $209$   &  $292.6$ &  $5/2$ &  $8.4$     \\
			CoO\cite{CoO}  &  $289$\cite{CoO}      &  $201$   &  $335$ &  $3/2$ &  $15.9$     \\
			CoPS$_3$\cite{CoPS3}  &  $120$\cite{CoPS3}      &  $61$   &  $101.7$ &  $3/2$ &  $15.3$     \\
			Cr$_2$O$_3$\cite{Cr2O3}  &  $308$\cite{Cr2O3}      &  $194$   &  $323.3$ &  $3/2$ &  $5.0$     \\
			Cr$_2$TeO$_6$\cite{Cr2TeO6}  &  $93$\cite{Cr2TeO6}      &  $53$   &  $88.3$ &  $3/2$ &  $5.0$     \\
			Cr$_2$WO$_6$\cite{Cr2WO6}  &  $45$\cite{Cr2WO6}      &  $27$   &  $45$ &  $3/2$ &  $0.0$     \\
			CrBr$_3$\cite{CrBr3}  &  $32$\cite{CrBr3}      &  $20$   &  $33.3$ &  $3/2$ &  $4.2$     \\
			CrCl$_2$\cite{CrCl2}  &  $20$\cite{CrCl2}      &  $15$   &  $22.5$ &  $2$ &  $12.5$     \\
			CrCl$_3$\cite{CrCl3}  &  $14$\cite{CrCl3}      &  $10$   &  $16.6$ &  $3/2$ &  $18.6$     \\
			CrI$_3$\cite{CrI3}  &  $61$\cite{CrI3}      &  $39$   &  $65$ &  $3/2$ &  $6.6$     \\
			CuFeO$_2$\cite{CuFeO2}  &  $11$\cite{CuFeO2}      &  $7.5$   &  $10.5$ &  $5/2$ &  $4.5$     \\
			CuO\cite{CuO}  &  $213$\cite{CuO}      &  $72$   &  $216$ &  $1/2$ &  $1.4$     \\
			EuO\cite{EuO}  &  $69.15$\cite{EuO}      &  $52$   &  $66.9$ &  $7/2$ &  $3.3$     \\
			EuS\cite{EuS}  &  $16.57$\cite{EuS}      &  $13$   &  $16.7$ &  $7/2$ &  $0.8$     \\
			Fe$_2$O$_3$\cite{Fe2O3}  &  $960$\cite{Fe2O3}      &  $730$   &  $1022$ &  $5/2$ &  $6.4$     \\
			FeCl$_2$\cite{FeCl2}  &  $23.55$\cite{FeCl2}      &  $11.5$   &  $23$ &  $1$ &  $2.3$     \\
			FeF$_2$\cite{FeF2}  &  $78.4$\cite{FeF2}      &  $55$   &  $82.5$ &  $2$ &  $5.2$     \\
			FeO\cite{FeO}  &  $198$\cite{FeO}      &  $148$   &  $222$ &  $2$ &  $12.1$     \\
			FePS$_3$\cite{FePS3}  &  $120$\cite{FePS3}      &  $91$   &  $136.5$ &  $2$ &  $13.8$     \\
			FePSe$_3$\cite{FePSe3}  &  $110$\cite{FePSe3}      &  $75$   &  $112.5$ &  $2$ &  $2.3$     \\
			FePt$_3$\cite{FePt3}  &  $170$\cite{FePt3}      &  $104$   &  $167.0$ &  $1.65$ &  $1.8$     \\
			HoFeO$_3$\cite{HoFeO3}  &  $647$\cite{HoFeO3}      &  $456$   &  $638.4$ &  $5/2$ &  $1.3$     \\
			KCoF$_3$\cite{KCoF3}  &  $114$\cite{KCoF3_T}      &  $33$   &  $99$ &  $1/2$ &  $13.2$     \\
			KCuF$_3$\cite{KCuF3}  &  $39$\cite{KCuF3}      &  $15$   &  $45$ &  $1/2$ &  $15.4$     \\
			La$_2$CoO$_4$\cite{La2CoO4}  &  $275$\cite{La2CoO4}      &  $161$   &  $268.3$ &  $3/2$ &  $2.4$     \\
			La$_2$NiO$_4$\cite{La2NiO4}  &  $328$\cite{La2NiO4}      &  $220$   &  $440$ &  $1$ &  $34.1$     \\
			LaFeO$_3$\cite{LaFe(V)O3}  &  $738$\cite{LaFeO3_T}      &  $504$   &  $705.6$ &  $5/2$ &  $4.4$     \\
					\end{tabular}
		\end{minipage}
	\hfill
\begin{minipage}[t]{0.42\textwidth}
%\centering
\begin{tabular}[t]{lccccc}
\hline
&  \makecell{$T_{\mathrm{Exp.}}$(K)} &  \makecell{$T_{\mathrm{MC}}$(K)}  & \makecell{$T_{\mathrm{MC}}^*$(K)}&  \makecell{S} &  \makecell{Error(\%)}\\
\hline
			LaMnO$_3$\cite{LaMnO3}  &  $139.5$\cite{LaMnO3}      &  $101$   &  $151.5$ &  $2$ &  $8.6$     \\
			LaVO$_3$\cite{LaFe(V)O3}  &  $150$\cite{LaVO3_T}      &  $70$   &  $140$ &  $1$ &  $6.7$     \\
			LiFePO$_4$\cite{LiFePO4}  &  $50$\cite{LiFePO4}      &  $41$   &  $61.5$ &  $2$ &  $23$     \\
			LuMnO$_3$\cite{LuMnO3}  &  $87.5$\cite{LuMnO3}      &  $71$   &  $106.5$ &  $2$ &  $21.7$     \\
			MnBi\cite{MnBi}  &  $630$\cite{MnBi}      &  $394$   &  $591$ &  $2$ &  $6.2$     \\
			MnBi$_2$Te$_4$\cite{MnBi2Te4}  &  $24$\cite{MnBi2Te4}      &  $16$   &  $22.4$ &  $5/2$ &  $6.7$     \\
			MnF$_2$\cite{MnF2}  &  $67$\cite{MnF2}      &  $48$   &  $67.2$ &  $5/2$ &  $0.3$     \\
			MnO\cite{MnO}  &  $117$\cite{MnO}      &  $79$   &  $110.6$ &  $5/2$ &  $5.5$     \\
			MnPS$_3$\cite{MnPS3}  &  $78$\cite{MnPS3}      &  $61$   &  $85.4$ &  $5/2$ &  $9.5$     \\
			MnPSe$_3$\cite{MnPSe3}  &  $74$\cite{MnPSe3}      &  $52$   &  $72.8$ &  $5/2$ &  $1.6$     \\
			MnS\cite{MnS}  &  $152$\cite{MnS}      &  $105$   &  $147$ &  $5/2$ &  $3.3$     \\
			MnS$_2$\cite{MnS2}  &  $48$\cite{MnS2}      &  $25$   &  $35$ &  $5/2$ &  $27.1$     \\
			MnTe\cite{MnTe}  &  $310$\cite{MnTe}      &  $251$   &  $351.4$ &  $5/2$ &  $13.4$     \\
			MnTiO$_3$\cite{MnTiO3}  &  $65$\cite{MnTiO3}      &  $45$   &  $63$ &  $5/2$ &  $3.1$     \\
			MnWO$_4$\cite{MnWO4}  &  $13.5$\cite{MnWO4}      &  $9$   &  $12.6$ &  $5/2$ &  $6.7$     \\
			NdFeO$_3$\cite{NdFeO3}  &  $693$\cite{NdFeO3}      &  $463$   &  $648.2$ &  $5/2$ &  $6.5$     \\
			NiBr$_2$\cite{NiBr2}  &  $52$\cite{NiBr2}      &  $28$   &  $56$ &  $1$ &  $7.7$     \\
			NiCl$_2$\cite{NiCl2}  &  $52.3$\cite{NiCl2}      &  $36$   &  $72$ &  $1$ &  $37.7$     \\
			NiF$_2$\cite{NiF2}  &  $73.2$\cite{NiF2}      &  $42$   &  $84$ &  $1$ &  $14.8$     \\
			NiO\cite{NiO}  &  $523$\cite{NiO}      &  $310$   &  $620$ &  $1$ &  $18.5$     \\
			PrFeO$_3$\cite{PrFeO3}  &  $707$\cite{PrFeO3}      &  $471$   &  $659.4$ &  $5/2$ &  $6.7$     \\
			Rb$_2$MnF$_4$\cite{Rb2MnF4}  &  $38.4$\cite{Rb2MnF4}      &  $33$   &  $46.2$ &  $5/2$ &  $20.3$     \\
			RbMnCl$_3$\cite{RbMnCl3}  &  $94$\cite{RbMnCl3}      &  $69$   &  $96.6$ &  $5/2$ &  $2.8$     \\
			RbMnF$_3$\cite{RbMnF3}  &  $82.6$\cite{RbMnF3}      &  $54$   &  $75.6$ &  $5/2$ &  $8.5$     \\
			RbNiF$_3$\cite{RbNiF3}  &  $133$\cite{RbNiF3}      &  $79$   &  $158$ &  $1$ &  $18.8$     \\
			Sr$_2$NiWO$_6$\cite{Sr2NiWO6}  &  $54$\cite{Sr2NiWO6}      &  $26$   &  $52$ &  $1$ &  $3.7$     \\
			SrMnBi$_2$\cite{SrMnBi2}  &  $290$\cite{SrMnBi2}      &  $219$   &  $306.6$ &  $5/2$ &  $5.7$     \\
			SrMnSb$_2$\cite{SrMnSb2}  &  $297$\cite{SrMnSb2_T}      &  $189$   &  $283.5$ &  $2$ &  $4.5$     \\
			TbFeO$_3$\cite{TbFeO3}  &  $681$\cite{TbFeO3}      &  $456$   &  $638.4$ &  $5/2$ &  $6.2$     \\
			TmFeO$_3$\cite{TmFeO3}  &  $635$\cite{TmFeO3_T}      &  $442$   &  $618.8$ &  $5/2$ &  $2.6$     \\
			YbFeO$_3$\cite{YbFeO3}  &  $627$\cite{YbFeO3_T}      &  $440$   &  $616$ &  $5/2$ &  $1.8$     \\
			YbMnBi$_2$\cite{YbMnBi2}  &  $290$\cite{YbMnBi2}      &  $203$   &  $284.2$ &  $5/2$ &  $2.0$     \\
			YbMnSb$_2$\cite{YbMnSb2}  &  $345$\cite{YbMnSb2}      &  $266$   &  $372.4$ &  $5/2$ &  $7.9$     \\
			YFeO$_3$\cite{YFeO3}  &  $644.5$\cite{YFeO3_T}      &  $445$   &  $623$ &  $5/2$ &  $3.3$     \\
			YMnO$_3$\cite{YMnO3}  &  $70$\cite{YMnO3}      &  $40$   &  $60$ &  $2$ &  $14.3$     \\
			YVO$_3$\cite{YVO3_T}  &  $118$\cite{YVO3_T}      &  $64$   &  $128$ &  $1$ &  $8.5$     \\
		\end{tabular}
	\end{minipage}
    %\hfill
    \hspace*{1.2cm}
\end{table}

\end{widetext}
\bibliography{refs.bib}

\end{document}